\documentclass{article}
\usepackage[english]{babel}
\usepackage{amssymb}
\usepackage{amsfonts}
\usepackage{amsmath}
\usepackage{graphicx}
\usepackage{epsfig}
\usepackage{cite}
\newcommand{\be}{\begin{equation}} \newcommand{\ee}{\end{equation}}
\newcommand{\bea}{\begin{eqnarray}} \newcommand{\eea}{\end{eqnarray}}
\newcommand{\beann}{\begin{eqnarray*}}  \newcommand{\eeann}{\end{eqnarray*}}
\newcommand{\bfig}{\begin{figure}} \newcommand{\efig}{\end{figure}}
\newcommand{\ba}{\begin{array}} \newcommand{\ea}{\end{array}}
\newcommand{\bcen}{\begin{center}} \newcommand{\ecen}{\end{center}}
\newcommand{\btab}{\begin{tabular}} \newcommand{\etab}{\end{tabular}}

\begin{document}
\title{\bf \huge Non-linear, Finite Frequency Quantum Critical Transport from AdS/CFT}

\author
 {%
   Andreas Karch$^1$
and
S. L. Sondhi$^2$
\vspace{0.4cm}
\\
\small $^1$Department of Physics, University of Washington, Seattle, WA 98195-1560, USA\\
\small $^2$Department of Physics, Joseph Henry Laboratories,
Princeton University, Princeton, NJ 09544, USA\\
\small ~\,E-mail: \tt{ karch@phys.washington.edu, sondhi@princeton.edu}
}

\maketitle

\centerline{\bf Abstract:}
Transport at a quantum critical point depends sensitively on the relative magnitudes of temperature, frequency and electric field. Here we used the gauge/gravity correspondence to
compute the full temperature and, generally nonlinear, electric field dependence of the electrical conductivity for some model critical theories. In the special case of 2+1 dimensions we are also able to find the full time-dependent response of the system to an arbitrary time dependent external electric field. The response of the system is instantaneous, implying a frequency independent conductivity. We describe a mechanism that rationalizes the instantaneous response.

\section{Introduction}

Quantum critical points continue to be objects of intense attention for reasons ranging from their intrinsic elegance to their potential for economically organizing our understanding of the phase
diagrams of various strongly correlated condensed matter systems \cite{sachdev}. For example,
unusual power laws in electrical transport have often been attributed to the proximity of the
system to various quantum critical points, most notably in the case of the cuprates and the
heavy fermion compounds.

By definition, quantum critical points lead to singular behavior in various physical quantities
and a key implication of this is that one cannot understand their transport properties in terms
of a set of transport coefficients, such as an electrical conductivity. Instead one needs to
compute the full functional dependence of currents on fields allowing for variation of the frequency $\omega$, temperature $T$ and amplitude $E$ (for the case of electrical transport that we consider in this paper). This requirement arises as the various limits of small $\omega$, $T$ and $E$ do
not commute.

For example, it was pointed out in \cite{green-2005} that the current $j(\omega, E)$
established in response to a finite frequency and finite amplitude electric field should
in general be expected to be a non-trivial function of the the dimensionless
combination $\omega/E^{z/(1+z)}$ for a quantum critical point with dynamic scaling exponent
$z$. This has the implication that a linear response computation of the DC transport,
which takes the limit $E \rightarrow 0$ before $\omega \rightarrow 0$ should be expected
to disagree with the answer obtained when the limits are reversed. Indeed, in the case
studied in \cite{green-2005} this was shown explicitly. This discussion parallels
a similar competition between the $\omega \rightarrow 0$ and
$T \rightarrow 0$ limits first described in \cite{damle-1997-56}.

Let us now state more precisely the challenge considered in this paper. Quite generally, scaling
arguments imply that the electrical current at a quantum critical point in $d_s$ spatial
dimensions characterized by a correlation length exponent $\nu$ and dynamic scaling exponent $z$
takes the form:
\be
\label{genscalingform}
 j(\omega, T, E) = E^{d_s-1 + z \over 1 + z} \, Y(E/T^{1 + 1/z},\, \omega/T) \ .
\ee
Thus the broad aim of a detailed theory of a given critical point is, in addition to
computing $\nu$ and $z$, to compute the function $Y$ and that is what we address in this
paper. Let us note that this is a non-trivial
task. In principle, in the limit $E^{1 + 1/z} \ll \omega, T$ we can resort to linear response
theory and the Kubo formula. In practice, as noted in \cite{damle-1997-56}, this is not
always straightforward. However, matters are even trickier when we move away from this limit
of small $E$ for now the response is generically nonlinear in $E$ with a fractional power
and one cannot resort to correlation function computations at
all \cite{green-2005, dalidovich-2004}. For example, in the limit $\omega, T \ll E$ scaling
predicts the algebraic form
\be
\label{scaling}
j \propto E^{\frac{d_s-1+z}{1+z}}.
\ee

Two previous computations that have accessed this nonlinear regime (although not the full function $Y$) by means of a transport equation \cite{green-2005} and a clever choice of a dissipative,
Gaussian, critical point \cite{dalidovich-2004}. In this paper we bring newer tools to bear
on this, specifically the AdS/CFT correspondence.

For a large class of quantum critical points with a dual gravitational description in terms of the AdS/CFT correspondence the finite $T$ non-linear DC  ($\omega=0$) conductivity can be readily
obtained.
We will review this construction in the next section. However, AdS/CFT allows
us to extract much more information about these critical points than just their DC transport.
Specifically, in the special case of 2+1 dimensional quantum critical points with relativistic
scaling ($z=1$) we are able to construct a full time-dependent solution that describes
the response to an electric field $E(t)$ with arbitrary time dependence, both at zero and finite temperature, thus obtaining the {\it full} function $Y$ which is, remarkably, simply a constant.
This builds in the physics that the resulting current responds to the external field instantaneously and that is simply linear in $E$ irrespective of the amplitude. Equivalently, the conductivity
$ j(\omega, T, E)/E = \sigma(\omega/T,\omega/\sqrt{E})$ is completely independent of the two variables it could potentially depend on. This finding nicely extends the result of \cite{Herzog:2007ij} where it was shown that, in a slightly different theory with gravity dual, the conductivity in the absence of electric fields was independent of frequency and thus of the first argument. In that work it was argued that this independence was a result of electric/magnetic duality in the gravitational description. Our result indicates that this $\omega$ independence is much more general in theories with a gravitational dual.

Before proceeding to the details some comments on the potential origins and validity
of nonlinear response are in order. Note first that according to the work energy theorem the total energy of the system in a putative transport steady state increases at a constant rate $E j$ due to Joule heating. It was argued in \cite{green-2005} that for small electric field $E$ this Ohmic heating can be neglected, as it is of one higher power in the electric field than the current. As $E$ is a dimensionful quantity what this really means is that Ohmic losses are negligible up to times of the order $t_E =E^{-1/(z+1)}$. For a time dependent electric field that is ramped up from zero to some finite value $E$ over a time period $\tau$ one would expect that the system should settle to a steady state over a time scale set by $\tau$ and then remain in this steady state up to time $t_E$ by which point Ohmic heating will have had enough time to build up a non-negligible finite temperature. As long as $\tau \ll t_E$ there should be a window in which the scaling solution (\ref{scaling}) applies.

The microscopic picture proposed in \cite{green-2005} as underlying this steady state solution was a competition between Schwinger pair production (which increases the available densities of charge carrier pairs) and scattering losses. Taking these two processes into account one expects the rate of change of the current to be given by (focusing on the relativistic case $z=1$)
\be
\label{micromodel}
\frac{d j}{dt} = a E^{(d_s +1)/2} - b \sqrt{E} j
\ee
The first term is the Schwinger process, the second term is the scattering loss. The $E$ dependence of the relaxation rate is once more determined by scaling alone\footnote{For working out the steady state response to a field it does not matter whether we use the field itself to set the scale of the scattering rate as we do here or use the current. At zero field the latter is clearly necessary.
Later we will use (\ref{micromodel}) in the time domain but again it will not matter which form
we use.}.
The numbers $a$ and $b$ should be of order 1 in a generic theory; however, in the examples we will be interested in where the field theory has some large intrinsic dimensionless parameters (such as $N_c$, the number of colors and $\lambda$, the 't Hooft coupling in our main example of a supersymmetric gauge theory), $a$ and $b$ can of course scale with these. The resulting current in the steady state is $j=a/b E^{d_s/2}$. Later we will make sense of our AdS/CFT results in the
light of these ideas.

Turning to the balance of this paper, Section 2 contains a review of the method of refs \cite{Karch:2007pd,O'Bannon:2007in} to calculate DC conductivities in a large class of theories with gravitational dual. We will show that these systems do indeed exhibit the scaling solutions predicted by \cite{green-2005}. In Section 3 we then focus on the 2+1 dimensional system and construct the full solution in response to a time dependent external electric field. From this we extract the conductivity. We also discuss implications for the validity of the simple microscopic model (\ref{micromodel}). In section 4 we discuss the action of S-duality in the 2+1 dimensional case and its connection to the results we obtained. We conclude in section 5.

\section{DC conductivities via AdS/CFT}
\subsection{Basic Philosophy}

The basic objects that can be calculated in AdS/CFT are correlation
functions of gauge invariant operators. Typically one uses
this to extract conductivities as a low-frequency limit
of a two point function (which is just governed by
the bulk propagator) and using a Kubo formula
with all the limitations described in the Introduction.
An alternative strategy developed
in \cite{Karch:2007pd,O'Bannon:2007in} is
to turn on $E$ directly as an external source. In this deformed
theory one then simply needs to calculate a one-point
function, $\langle j_x \rangle$. For this to be possible
the system needs to settle to a stationary state.
For the ``probe brane systems" considered in \cite{Karch:2007pd,O'Bannon:2007in}, the energy and momentum loss rates
are of order $N$, whereas background energy densities are of order $N^2$.
So for times that are not parametrically large in $N$ the background
glue acts as an effective heat and momentum sink that lets the
system come to a stationary state. In this sense the situation is even better than in the generic quantum critical system: the loss rates are negligible not just to time scales of order $t_E$ but actually all the way to $N \, t_E$.

To calculate a one-point function all one needs to do is
solve for the near boundary behavior of all the fields in the bulk.
Close to the boundary bulk fields fall off as a power of the radial
coordinate $r$ (where $r \rightarrow \infty$ is the boundary) and
so these powers can be read off from linearized equations of motion.
For AdS$_{d+1}$ Maxwell's equations tell us that gauge fields have
two modes, one constant and one that falls of as $r^{2-d}$. The former
is non-normalizable. Its coefficient corresponds to the source
in the field theory. The variation of the on-shell action
with respect to the source gives the expectation values. Those
depend on the subleading coefficient. For a gauge field,
this can easily be expressed in terms of the field strength.
Let us split the coordinates into the field theory
coordinates $x^{\mu}$ and the radial coordinate $r$.
$F_{\mu \nu}$ asymptotically approaches a constant whose
value directly corresponds to $F_{\mu \nu}$ in the field theory.
So for the DC conductivity we are interested in a solution
to the bulk equations of motion where $F_{xt}$ asymptotically
reaches $E$, the boundary electric field. To turn on a magnetic
field in addition one also needs to turn on $F_{xy}=B$. On the
other hand, $F_{r \mu}$ asymptotically falls off
as $\langle j^{\mu} \rangle r^{1-d}$. So for the conductivity
all we need to find is $F_{xr}$ to extract $j_x$. In the more general setup of \cite{O'Bannon:2007in} one
has $E$ and $B$ turned
on as well as $F_{xr}$, $F_{rt}$ (which signals a finite density,
$F_{rt} \sim \langle j_t \rangle r^{1-d}$) and then also $F_{ry}$ (since
with both magnetic field and density non-zero one also
gets a Hall current). The most
general system studied in the literature \cite{Ammon:2009jt} allows in addition for a non-trivial $\vec{E} \cdot {\vec B}$.

For the simplest DC conductivities (no density, no magnetic field) we
are looking for a solution to the equations of motion of a gauge field
in AdS$_{d+1}$ where $F_{xt} = E$ and $F_{xr}$ are the only
non-vanishing components of the field strength. For a given $E$,
the leading near boundary behavior of $F_{xr}$ gives the current and
hence the conductivity. But how does one fix $F_{xr}$ in terms of $E$?
If the field in the bulk is obeying standard Maxwell equations,
$F_{xt}=E$, $F_{xr} \sim \langle j_x \rangle r^{1-d}$ is a solution
for any value of $E$ and $\langle j_x \rangle$. We need one more boundary
condition to relate the two. As this is supposed to be a result of the non-trivial IR
dynamics this boundary condition has to be a regularity condition imposed in the interior of AdS. For Maxwell alone one would for example
have to demand $\langle j_x \rangle =0$ for any $E$ as otherwise
$F_{xr}$ blows up as $r \rightarrow 0$. At finite temperature
the minimal value of $r$ is the horizon radius and all solutions
seem to be fine. Either way, this is clearly not the right physics.
Maxwell alone does not describe a consistent dual field theory. As the field
strength grows in the interior one needs to know the full non-linear
equation obeyed by the gauge field. Identifying the right IR boundary
conditions depends on the details of the setup.

For the class of theories that are described by ``probe branes" \cite{Karch:2000gx,Karch:2002sh} (we'll give examples of this type of construction below) the full non-linear action is given by the Dirac-Born-Infeld (DBI) action
\be
\label{DBI}
S = - {\cal N} \int dr dx^d \sqrt{- \det(g+F)} =
- {\cal N} \int dr dx^d \sqrt{-g} \sqrt{1 + F^2}.
\ee
For the last step we assumed that $F \wedge F =0$, otherwise
additional terms are present when expanding the determinant of $g+F$ in
powers of $F$. ${\cal N}$ is a normalization
constant that is completely determined
by the tension of the brane and the internal geometry; it is proportional
to $N_c$, the number of colors in the gauge theory whose gravitational dual we are studying.  The structure of the DBI action gives solutions
for the field strength that involve square roots; a generic solution
that is real near the boundary will turn complex at some radial coordinate.

The dynamics underlying this feature of the DBI action are
related to Schwinger pair production, this time in the bulk. The DBI action
is an effective action describing the dynamics of open strings ending on the D-brane. In the limit of large string tension all the excited string modes can be integrated out and we obtain an effective action for the
zero modes of the string, which are the gauge fields living on the D-brane as well as the scalar fields describing its transverse position. The endpoints of the string are charged under the gauge field living on the brane. In the presence of a small background electric field the tension in the string overcomes the force that wants to pull
its endpoints apart. As the
latter is independent of the applied electric field, there exists a critical electric field
beyond which strings ending on the brane become unstable. This is the point at which the DBI action becomes complex. If we impose such an electric field as an initial condition, the system is unstable.

But even in the presence of a background $E$ field real solutions can
be found if one turns on $F_{xr}$ (corresponding to a current $\langle j_x \rangle$ in the dual field theory)
in addition to $E$.
It turns out that, at least for time-independent $E$, there exists
a unique value of $\langle j_x \rangle$ for which the solution
remains real. This is the IR boundary condition that fixes
the conductivity. In the time dependent case we are going to exhibit
an analytic solution that seems to be manifestly free of pathologies.
Presumably in this case this is also the unique choice.

As an example, let's review the conductivity of the D3/D7 system
corresponding to a 3+1 dimensional gauge theory with order $N_c^2$ gluons coupled
to order $N_c$ matter fields, bosonic and fermionic.
The metric we are using in the bulk is that of an AdS black hole
\be ds^2 = -h(r) dt^2 + dr^2/h(r) + r^2 d \vec{x}^2 \ee
where $h(r)= r^2 -r_h^4/r^2$, and the horizon radius $r_h$ sets
the temperature as $r_h = \pi T$. Working at zero $B$, zero density
the only field we need to turn on is $A_x =- E t - H(r)$,
so that $F_{xt} = E$ and $F_{xr}= H'(r)$. The action in this case becomes
\be S = {\cal N} \int dr dt r^3 \sqrt{1 - F_{xt}^2/(h r^2) + h/r^2 F_{xr}^2}.\ee
It is easy to see that constant $F_{xt}=E$ solves the $t$ component of the equations of motion
that follow from this action. The only non-trivial equation comes from the $x$ component of the equations of motion.
As
$A_x$ appears in the action only via its derivatives,
we can integrate the equations of motion and get
\be
\label{d7sol}
 H'(r) =  \langle j_x \rangle \sqrt{  \frac{ ( r^2 - E^2/h)}{
{\cal N}^2 h^2 r^4 -  h \langle j_x \rangle^2}}
\ee
where at this stage $\langle j_x \rangle$ simply is an undetermined
integration constant. Recall that $h(r) \sim r^2$ at large $r$ and decreases
monotonically as we decrease $r$ with $h(r)=0$ at $r=r_h$. So
both numerator and denominator under the square root change sign
at some value of $r$. The only way for $H'$ to stay real
is if numerator and denominator change sign at the {\it same} value
of $r$.
The numerator changes sign at
a value $r_*$ that is set solely by the external field,
$ r_*^2 h(r_*) = E^2$
and hence $r_*^4 = E^2 + (\pi T)^4$.
For small electric field $r_*=r_h$ and the corresponding transport phenomena
are completely dominated by the horizon, which is supposed to roughly
encode the physics of the longest wavelength fluctuations. As we increase $E$,
$r_*$ moves outwards and starts sampling the full geometry. So
the non-linear transport depends on the details of the geometry.
The current $\langle j_x \rangle$ is now completely determined by demanding
that the zero of the denominator is at the same point $r_*$, that
is ${\cal N}^2 h(r_*) r_*^4 = \langle j_x \rangle^2$ and so
\be
\langle j_x \rangle = \frac{N_f N_c}{ 4 \pi^2} \left ( E^2 + (\pi T)^4
\right )^{1/4} E.
\ee
where in the last step we plugged in the value of ${\cal N}$
corresponding to a D7 on AdS$_5$ $\times$ $S^3$.
In the notation introduced in (\ref{genscalingform}) this implies the identification,
\be
Y(x,0) = \frac{N_f N_c}{ 4 \pi^2} \left[ 1 + \left(\frac{\pi}{x}\right)^2 \right]^{1/4}
\ee
To compare to
the answer quoted in \cite{Karch:2007pd} one needs to further
rescale $E$ and $\langle j_x \rangle$ by $\frac{\sqrt{\lambda}}{2 \pi}$. The
normalization used in \cite{Karch:2007pd} is the natural one where
a single charge carrier has charge 1. The normalization here avoids
prefactors of $\frac{\sqrt{\lambda}}{2 \pi}$ in the DBI action
by assigning charge
$\frac{2 \pi}{\sqrt{\lambda}}$ to the basic charge carriers.

This result already
shows the crucial points about the limitations of linearized response: the linearized response
answer can simply be obtained
by linearizing $\langle j_x \rangle$
in $E$, so $\sigma = \frac{N_f N_c}{4 \pi} T$. This vanishes in the zero temperature limit.
Going first to zero temperature on the other hand gives
us $\sigma = \frac{N_f N_c}{4 \pi^2} \sqrt{E}$. Similar expressions
can be worked out at finite density and magnetic field and they basically
all exhibit this feature. One
can also obtain the conductivity from the $\omega \rightarrow 0$ limit of a frequency dependent 2-pt function via a Kubo formula, as was
explicitly done for this system in \cite{Mas:2008jz}. This calculation is in perfect agreement with the $E=0$ answer we obtained above (and therefore also fails to capture the finite answer one gets by setting $\omega=T=0$ at finite $E$).
In order to get a frequency dependent conductivity from our approach of calculating 1-pt functions in the presence of a finite electric field we need to turn on a time-dependent electric field. We will do so in the case
of a 2+1 dimensional
quantum critical point below.

\subsection{Generalizations}
The calculation described above can easily be generalized to other field theories that have a dual in terms of gravity coupled to probe branes. By changing the probe brane we can couple matter localized on $d_s+1$ dimensional defects to the same 3+1d conformal field theory living on the D3 branes. One can also change the dimensionality of the glue theory, but typically SYM is not a conformal theory in spacetime dimensions other than 3+1, so for applications to the physics of quantum critical points we will only concern ourselves with conformal defects coupled to 3+1 dimensional glue.

For a probe brane in ${\cal N}=4$ SYM giving rise to a defect
with massless matter living on
$d_s$ spatial dimensions, one finds a DC conductivity
\be
\sigma = \sigma_0 (E^2 + (\pi T)^4)^{(d_s-2)/4}
\ee
where $\sigma_0$ is a pure number that can be calculated in terms
of the details of the probe (its tension and the volume
of the internal space it wraps) \cite{Karch:2007pd}.
Clearly for any $d_s \neq 2$ one fails to find the correct non-linear conductivity when working with
zero background electric field and then taking the limit of small $T$, whereas a zero $T$ analysis with
finite $E$ does indeed give the predicted scaling result.

This calculation can also be generalized to Lifshitz backgrounds,
that is metrics that have a scale invariance with a dynamical
critical exponent $z$ different from 1. The symmetries
require the metric to take the form \cite{Kachru:2008yh}
\be
ds^2 = -r^{2 z} dt^2 + r^2 d \vec{x}^2 + dr^2/r^2.
\ee
One can study the same DBI action on this background to obtain
a conductivity via Ohm's law\footnote{Some aspects of the DC conductivity in these backgrounds as obtained by a probe brane analysis similar to the one performed here were also presented in \cite{Hartnoll:2009ns,Ammon:2010eq,Fadafan:2009an}.}.
As in the D3/D7 example we discussed above, reality of the solution
imposes a boundary condition at a special radial coordinate $r_*$ given in
terms of the metric components $g_{tt}$ and $g_{xx}$ by
\be
g_{xx} |g_{tt}| = E^2
\ee
or in other words $r_*^{1+z} = E$. In \cite{Karch:2007pd} the conductivity was
shown to be
given by
\be
\sigma = \sigma_0 g_{xx}^{(d_s-2)/2}
\ee
for any diagonal metric, where the metric component once
more has to be evaluated at $r_*$. This reproduces
\be
\sigma = \sigma_0 E^{\frac{d_s-2}{1+z}}
\ee
as predicted by the scaling analysis of
\cite{green-2005}.

\subsection{Conductivity of the D3/D5 system}

As we have seen above,
the pattern of vanishing conductivities when working with $E=0$, $T \rightarrow 0$ missing the correct
 zero temperature answer is generic with one exception: 2+1 dimensional defects. In that case there is no non-linear dependence of the conductivity on the electric field and the limits of zero T and zero $E$ commute. In fact, the conductivity is completely independent of both. The easiest example of a 2+1 dimensional defect coupled to the D3 brane glue is given by a D5 brane probe on AdS$_4$ $\times$ $S^2$, as this theory preserves half of the supersymmetries of the glue and so stability is guaranteed.
In this case (at zero density, zero magnetic field and for massless flavors) the
conductivity $\sigma$ is given by
\be \sigma = \frac{2 N_f N_c}{\pi \sqrt{\lambda} }.\ee
This formula is true at all values of $E$ and $T$ and is independent of the dimensionless ratio $E/T^2$. Scale invariance alone would allow a non-trivial dependence on this ratio, so the independence seems to be a special property of
field theories with a probe brane gravitational dual.

A very similar phenomenon in theories with holographic dual has been observed before in \cite{Herzog:2007ij}. There it was shown that in 2+1 dimensional theories
with gravitational dual in linearized response (that is at $E=0$)
$\sigma$, which should be a non-trivial function of $\omega/T$, is
in fact completely frequency independent.
Together with the result on the $E/T^2$ independence of the DC conductivity this strongly suggest that $\sigma(\omega,T,E)$ is
actually constant. In order explore this in our framework
we want to calculate the one-point function of $j_x$ not in response
to a constant electric field, but to a time varying electric field.
We will address this problem in the next section.
Before we embark on that calculation, let us note that with either a finite density of
charge carriers or a finite magnetic field turned on $\sigma$ is
no longer independent of $E$ and $T$ and the linearized response once more fails to yield the desired zero temperature answer.

\section{Time dependent solutions}

\subsection{General Strategy}

For a time-dependent electric field
one can still define a conductivity via Ohm's law
\be
\langle j_x (t) \rangle = \int d\tau \sigma(\tau) E(t-\tau).
\ee
In order to calculate this time dependent $\langle j_x (t) \rangle$
we have to redo the calculation of the worldvolume gauge
fields in the presence of a time dependent electric field.
That is the boundary condition in the UV of the AdS geometry
is that $F_{xt} = E(t)$ at large $r$. The 2nd condition
in order to uniquely specify the field strength is still expected to be
the requirement that the solution is real.

\subsection{Exact time dependent solution in 2+1 dimensions}

So far we have been using the metric in its standard black hole form
\be ds^2 = - h(r) dt^2 + \frac{dr^2}{h}  + r^2 d \vec{x}^2. \ee
To study time dependent solutions a better suited time coordinate
is the incoming $v$
\be t= v + f(r) \ee
where \be f'(r) = -1/h(r).\ee
At large $r$ we have $t=v$ and both correspond to the field theory
time.
Causality demands that information about the time
varying electric field can at best propagate inwards at the speed of light.
Lines of constant $v$ are spacetime points that are connected by
lightrays to a boundary event at $t=v$ and so it is natural to look
for solutions in the bulk whose time dependence is just governed by $v$,
In terms of $v$ the metric reads
\be ds^2 = 2 dv dr - h dv^2 + r^2 d \vec{x}^2. \ee
In this coordinate system the non-vanishing components
of the inverse metric are $g^{rv}=g^{vr}=1$, $g^{rr} = h$ and
$g^{xx}= 1/r^2$.

Denoting quantities in the $t,r$ coordinates with bars, we can work
out the effect of the change of coordinates on field strength in
order to connect to the dictionary we established relating $\bar{F}_{xt}$ and
$\bar{F}_{xr}$ to $E$ and $\langle j_x \rangle$
respectively \footnote{Explicitly, the change of variables gives
$$
F^{xv} = \bar{F}^{xt} + \bar{F}^{xr}/h, \,\,\,\,\,\, F^{xr} = \bar{F}^{xr}.
$$
and hence
$$ F_{xr} = \bar{F}_{xr} - \bar{F}_{xt}/h, \,\,\,\,\,
F_{xv} = \bar{F}_{xt}. $$
}.
The
solutions we are interested in have only $F_{xv}=E(v,r)$
turned on.
So $F^{xr} = E/r^2$, hence
$\bar{F}^{xr} = E/r^2$,  $\bar{F}^{xt} =
- E/(h r^2)$ and last but not least
\be \bar{F}_{xr} = E/h, \,\,\,\,\,
\bar{F}_{xt} = E. \ee
So $F_{xv} = E(v,r)$ alone is sufficient to encode both a background
electric field and a current proportional to $E$.
Alternatively, one can directly calculate the conductivity in the $v$, $r$ coordinate system from variation of the
action with respect to $F_{xr}$.

The Lagrangian for a D5 brane with these worldvolume fields turned on
is
\be
{\cal L} \sim \sqrt{-\det(g+{\cal F})} \equiv r^2 \bar{\cal L} = r^2
\sqrt{ r^{-2} (2 F_{xr} F_{xv} + h F_{xr}^2) + 1}.
\ee
The resulting equation of motion is
\be
\label{eom}
\partial_r \left ( \frac{ F_{xv}  + h  F_{xr} }{\bar{\cal L}} \right )
+ \partial_v  \left ( \frac{ F_{xr}  }{\bar{\cal L}} \right ) =0
\ee
together with the Bianchi identity
\be \partial_r F_{xv} = \partial_v F_{xr}. \ee
We can find one simple analytic solution
for any external forcing function $E(v)$ and temperature $T$:
\be \label{solution} F_{xr}=0, \,\,\,\,\, F_{xv} = E(v), \,\,\,\,\, \bar{\cal L}=1. \ee
In the field theory this corresponds to a state with $j(t) = \sigma E(t)$.
The solution is real and completely well behaved and so
presumably corresponds to the unique physical solution.

What we have thus established is that in 2+1d one can again
find $\langle j_x (t) \rangle$ analytically even in the presence of
a time dependent electric field with the result that
\be \langle j_x (t) \rangle = \sigma E(t) \ee
that is the current adjusts itself instantaneously to the time
varying electric field. In terms of the general $\sigma(\tau)$ this
means that $\sigma(\tau) \sim \delta(\tau)$ and so its Fourier
transform $\sigma(\omega)$ is $\omega$-independent.
So indeed $\sigma(\omega, T ,E)$ in
the case of a 2+1 dimensional quantum critical point with
a gravity plus probe brane dual is completely independent
of its arguments confirming our earlier suspicion based on the
independence of $E/T^2$ at $\omega=0$ and of $\omega/T$ and $E=0$.

Last but not least, one can calculate the energy
density deposited into the field theory using the techniques of \cite{Karch:2008uy} (that is one simply
calculates the stress tensor of the probe D5) and finds that the energy density $\epsilon$ is given by
\be \epsilon = \int dt j(t) E(t) \ee
as it should be.

\subsection{A microscopic mechanism}

Having worked out the conductivity in the holographic theory and having established that, for $d_s=2$, the conductivity is independent of $T$, $E$ and $\omega$, we can go back and wonder whether this feature can be understood, at zero temperature,
in terms of the microscopic mechanism of Schwinger pair production competing with scattering losses. For the case of $d_s=2$ our model equation reads
\be
\label{micromodel2}
\frac{d j}{dt} = a E^{3/2} - b \sqrt{E} j.
\ee
From our knowledge of the conductivity in the stationary state we find again the ratio of $a$ and $b$,
\be a/b = \sigma = \frac{2 N_f N_c}{\pi \sqrt{\lambda} }.\ee
To pin down the scaling of $a$ and $b$ with $N_c$ and $\lambda$ we need more information. Note that it is natural for the Schwinger process to scale with the number $N_f N_c$ of quarks that can run in the loop, so presumably in terms of the $N_c$ counting $a \sim N_c$ and hence $b \sim 1$. But we also need to understand how $a$ and $b$ scale with the other large parameter, the 't Hooft coupling $\lambda$. For what follows we will assume that $b$ itself is large. A plausible scaling is to take $b \sim \sqrt{\lambda}$ and hence $a \sim 1$ as far as the scaling with $\lambda$ is concerned. This would imply that unsupported currents decay with a time scale that scales as $t_E/\sqrt{\lambda}$. But as we will demonstrate momentarily, any parametrically large $b$ is consistent with the frequency independent conductivity we found.

We can solve the 1st order transport equation (\ref{micromodel2}) for general time dependent $E$:
\be
\label{integralsolution}
j(t) = a \int_{- \infty}^t \, dt' \, E(t')^{3/2} e^{- b K(t,t')}
\ee
where the kernel $K(t,t')$ is given by
\be
K(t,t') = \int_{t'}^t \sqrt{E(\tilde{t})} d\tilde{t}.
\ee
For large $b$ the integral in (\ref{integralsolution}) is dominated by the upper integration boundary as all earlier times are exponentially suppressed.
As always when an integral is dominated by an endpoint, we can evaluate the leading contribution from integrating by parts
\begin{eqnarray}
\nonumber
j(t) &=& \left . \frac{a}{-b \, (\partial_{t'} K(t,t'))} E(t')^{3/2} e^{-b K(t,t')} \right |_{- \infty}^t + {\cal O}(b^{-2}) \\
&=& \frac{a}{b} E(t)  + {\cal O}(b^{-2}) = \sigma E(t) + {\cal O}(\frac{\sigma}{b})\ ,
\label{integralsolutiontwo}
\end{eqnarray}
which then makes sense of the instantaneous response.

It is interesting to note that the considerations embodied in (\ref{micromodel2}) also imply that absent an electric field the current must relax to zero as $1/t^2$. [The reader
should revisit Footnote 1 if this is not clear]. Clearly, our result thus far implies that the
current instantaneously drops to zero once the field is switched off which is a trivial example of
this scaling. It is, however, interesting to search for but fail to find a constant current carrying
solution to get
at this result a different way that is consonant with the AdS/CFT ideas we have used in this paper.
In fact a potential such solution, for the case of zero temperature, has already been constructed in \cite{Karch:2007br}. That work was mostly concerned with the phase diagram at finite density, but it was noted that in the end the answers only depend on $\langle j_{\mu} j^{\mu} \rangle$ as it should be by boost invariance and so one can easily extend the calculation in \cite{Karch:2007br} to the case of a spatial current density. For a solution with a persistent current in the $x$ direction the bulk solution reads
\be
F_{xt} = \frac{ \langle j_x \rangle}{\sqrt{{\cal N}^2 r^4 - \langle j^2 \rangle}}.
\ee
When $\langle j^2 \rangle = -d^2 + \langle j_x^2 \rangle$ is negative, this is a perfectly regular solution. However for the case of interest with $d=0$ and constant $\langle j_x \rangle$ there is a singularity at a finite value of $r$. Clearly the persistent current is {\it not} a stable background.

\section{S-duality}

Above we have shown by explicit construction of a time-dependent solution that in 2+1 dimensional field theories with a probe brane dual the conductivity does not depend on either the ratio $\omega/T$ or $\omega^2/E$.
This result extends the earlier findings of \cite{Herzog:2007ij}, who showed that for the holographic dual of the M2 brane theory in the absence of an electric field the conductivity is independent of $\omega/T$. As this result was obtained via a standard Kubo formula, only the quadratic Maxwell part of the bulk gauge field action was important. It was argued in that work that the underlying reason for the $\omega$ independence was S-duality.

We found a similar frequency independence for a system governed by the DBI action. Remarkably, the DBI action
 is known to also be invariant under a non-linear generalization of S-duality \cite{Gibbons:1995cv,Tseytlin:1996it,Gaillard:1997rt}. In terms of $\vec{E}$, $\vec{B}$ and $\vec{D}=\delta {\cal L}/
\delta \vec{E}$ and $\vec{H} = -\delta {\cal L}/\delta \vec{B}$ its action is just the standard electro-magnetic
duality transformation of Maxwell's equations in matter. In this section we want to see to what extent S-duality is related to
the frequency independence in the system we analyzed.

It was argued in \cite{Witten:2003ya} that for any 2+1 dimensional conformal theory with a global $U(1)$ symmetry one can define a T-transformation and an S-transformation that takes the theory into a different conformal field theory (CFT) whose conductivities are related to the ones in the original CFT. For CFTs with a bulk dual these operations correspond to an electric-magnetic duality transformation.
The S-operator acts on the boundary field theory quantities as
\be
\label{sboundary}
j_x \rightarrow - \sigma_0 E_y, \,\,\,\,\, j_y \rightarrow - \sigma_0 E_x, \,\,\,\,\, E_y \rightarrow j_x/\sigma_0,
\,\,\,\,\, E_x \rightarrow j_y/\sigma_0, \,\,\,\,\, B\rightarrow -d/\sigma_0, \,\,\,\,\, d \rightarrow \sigma_0 B
\ee
These transformation rules can be derived most easily
from the bulk point of view.
The near boundary behavior and hence the transformation of the field theory
quantities is simply given by the standard Maxwell duality relations in AdS$_4$.
Near the boundary all components of the field strength approach a constant whereas the components of the inverse metric go to zero, so the linearized equations of motion derived from the Maxwell action hold.
The above transformation rules then follow recalling that $F_{\mu \nu}$ with both legs along the field theory directions approaches the corresponding field theory quantities, whereas the mixed components\footnote{For the purpose of discussing S-duality it is convenient to use a radial coordinate $u=1/r$.} $F_{u \mu}$ approach $\sigma_0 j_{\mu}$.

The S-operator can be applied irrespective of whether S-duality is a symmetry of the bulk action. In our case S-duality is indeed a symmetry of the classical equations of motion of the bulk theory so the theory generated by the S-transformation will have identical transport properties\footnote{The application of S-duality in the bulk is complicated by the fact that S-duality does not commute with the $g_s \rightarrow 0$, $4 \pi g_s N_c = \lambda$ fixed 't Hooft limit. S-duality takes the probe D5 brane into a probe NS5 brane. While both are described by a DBI action, the tension, which appears as a prefactor of the action, is of course different. As the classical equations of motion are independent of the prefactor of the action, the classical gravity + probe system has equations of motion that are invariant under S-duality and a subsequent constant shift in the dilaton which takes the theory back to weak coupling. So we expect the expressions for the conductivities to actually be invariant under S-duality.}. Let us demonstrate this in detail. According to the results of \cite{O'Bannon:2007in} the conductivities in the D3/D5 system in the presence of both electric and magnetic background fields as well as with a finite density of charge carriers are
\begin{eqnarray}
\nonumber
j_x &=& \frac{g_{xx}}{g^2_{xx} +B^2} \sqrt{\sigma^2 (g^2_{xx} + B^2) + d^2} E \\
\label{conductivities}
j_y &=& \frac{d B}{g^2_{xx} + B^2} E
\end{eqnarray}
where
\be g_{xx}^2 =\frac{1}{2} \left ( (\pi T)^4 + \vec{E}^2 - B^2 + \sqrt{(\vec{E}^2-B^2)^2 + 2(\pi T)^4 (\vec{E}^2+B^2)
+ (\pi T)^8} \right ) \ .
\ee

While we already know the transformation properties of the field theory quantities $j_{\mu}$, $B$ and $\vec{E}$,
the transformation rules for $g_{xx}$ need to still be determined. $g_{xx}$ stands for the value of the spatial components of the metric evaluated and a special radial location, the worldsheet horizon $u_*$. As the latter depends on the
full $u$ dependent profile of the bulk field strength and not just the near boundary behavior, one needs to take into account the S-duality of the full non-linear DBI action. From
the expressions derived in \cite{O'Bannon:2007in} it can easily be seen that $g_{xx}$ is invariant under the S-transformation. The location of the worldsheet horizon $u_*$ is determined in \cite{O'Bannon:2007in} as the simultaneous vanishing locus of 3 functions $\xi$, $\chi$ and $a$. $\xi$, $\chi$ and $a$ are all expressed in terms of $z$-independent boundary data (currents, electric fields, magnetic field, density) and so their transformation under S-duality is
 determined by the transformation properties of the boundary data (\ref{sboundary}). It is easy to see that under S-duality
\be \xi \rightarrow \chi/\sigma_0, \,\,\,\,\,  \chi \rightarrow \sigma_0 \xi, \,\,\,\,\, a \rightarrow a.
\ee
As $z_*$ is defined as the vanishing locus of $\xi$, $\chi$ and $a$ it remains invariant and so does
$g_{xx}(z_*)$.

Applying the transformation (\ref{sboundary}) to the expressions for the conductivities (\ref{conductivities}) one sees that in the dual field theory only $j_y$ is turned on, whereas the electric field now has both $E_x$ and $E_y$ components. The interesting physical quantities are the magnitude $j$ of the induced current as a function of the magnitude of the electric field, as well as the Hall angle, the angle between $\vec{j}$ and $\vec{E}$. After some algebra one finds that in the dual theory these two quantities are in fact {\it identical} to the original ones; electric field and resulting current in the S-dual system are completely equivalent to the original one up to an overall rotation in the $x$-$y$ plane. So the S-transformation generated a new theory which, by virtue of the S-duality of the DBI action, has identical transport properties.

However, unlike in the case of the M2 brane gauge theory studied in \cite{Herzog:2007ij}, S-duality does not impose any restriction on the conductivity of the original theory. The $\omega$ independence of the conductivity therefore seems to not be intrinsically connected to S-duality.
In fact,
had we taken any Lagrangian ${\cal L}(F^2)$ being an arbitrary function\footnote{\label{lfs} By a Lagrangian of the form ${\cal L}(F^2)$ we mean a Lagrangian that is $\sqrt{-g}$ times an arbitrary function $f(F^2)$ of $F^2=F_{\mu \nu}
g^{\mu \mu'} g^{\nu \nu'} F_{\mu' \nu'}$. The DBI itself is manifestly of that form with $f=\sqrt{1+F^2}$ as can be seen from eq. (\ref{DBI}). In completely parallel with the DBI equation of motion from eq. (\ref{eom}) the equation of motion in the general case is
\be
\partial_r \left [ f'(F^2) (F_{xv} + h F_{xr} ) \right ] + \partial_v \left [ f'(F^2) F_{xr} \right ] =0 
\ee
which is clearly still solved by $F_{xr}=0$ (and hence $F^2=0$) with $r$ independent $F_{xv}=E(v)$.
Of course, generic higher derivative interactions as they would be produced by quantum corrections would change the conclusions. In particular couplings of the form $R F^2$ in the action would spoil the solution (\ref{solution}). Also, if a background scalar field $\phi$ is turned on with a non-trivial profile, couplings of the form $F^2$ times a function of $\phi$ would ruin the time independent solution.} of $F^2$ in the same AdS or AdS black hole background, we would still get that the time dependent (\ref{solution}) solves the equations of motion for any time-dependent electric field on the boundary, with an instantaneous response of the current to the field. However S-duality would no longer be a symmetry of the bulk action. The instantaneous response therefore is a genuine property of any field theory with a classical holographic dual of the form spelled out in footnote \ref{lfs}  and does not seem to be deeply connected to S-duality.

\section{Conclusions}
We have analyzed electrical transport at the strongly coupled quantum critical points which have a dual holographic description in terms of probe branes. We find a stationary solution at an arbitrary temperature with a constant current and a generally non-linear dependence on the electric field.
This adds to the small \cite{green-2005,dalidovich-2004} catalog of quantum critical points where
the non-linear dependence has been explicitly computed. Of the previous computations, this present work is in the spirit of \cite{green-2005} where a regime of non-linear response is explicitly exhibited for a non-dissipative field theory in a large $N$ limit. Indeed, the explicit mechanism
involved in \cite{green-2005} provides a heuristic guide to our results. The non-linear dependence on $E$ is uniquely governed by scaling properties; the coefficient is finite and calculable in our models. Clearly this phenomenon is beyond the reach of linear response and so the application of Kubo formulas to the problem of transport at critical points misses this effect.

For the special case of 2+1 dimensions we were able to also find an analytic solution in the case of a time-dependent finite electric field, both at zero and finite temperature thus providing a
fairly general computation of the critical electrical transport. The resulting form is quite
unusual in that the current instantaneously adjusts to changes in the external field and is always
linear in the electric field. This extends earlier findings regarding the independence of the conductivities in 2+1 dimensional holographic models from $\omega/T$ at $E=0$ and from $E/T^2$ at $\omega=0$ to the full $\omega/T$, $\omega^2/E$ plane. While it was earlier argued that this frequency independence is a consequence of S-duality, in our case there seems to be no causal connection. While our system does exhibit a nice action of S-duality, this does not seem to force an $\omega$ independent conductivity all by itself. Furthermore, we are able to construct a large class of different 3+1d bulk theories that all exhibit frequency independent conductivities without being S-duality invariant.

Altogether, the results in this paper add to the growing set of results that illustrate the
power of the AdS/CFT correspondence to illuminate issues of principle in the study of quantum
critical phenomena.

\section{Acknowledgements}
AK would like to thank Andy O'Bannon for numerous helpful discussions. We are indebted to Ofer Aharony for useful email correspondence. Special thanks are owed to Ambika Sondhi and Mari
Torii-Karch for helping to initiate this collaboration. The work of AK was supported in part by the U.S. Department of Energy under Grant No.~DE-FG02-96ER40956 and SLS acknowledges support by the
NSF through grant DMR-1006608.

\bibliographystyle{JHEP}
\bibliography{realt}

\end{document}